\documentclass[a4paper,11pt]{article}
\usepackage{pos}

\title{Solving the Hubble tension à la Ellis \& Stoeger 1987}

\author*[a]{Jenny Wagner}

\affiliation[a]{Bahamas Advanced Study Institute and Conferences \\
4A Ocean Heights, Hill View Circle, Stella Maris, Long Island, The Bahamas}

\emailAdd{thegravitygrinch@gmail.com}

\abstract{The discrepancy between the value of the Hubble constant $H_0$ in the late, local universe and the one obtained from the Planck collaboration representing an all-sky value for the early universe reached the 5-$\sigma$ level. 
Approaches to alleviate the tension contain a wide range of ansatzes: increasing uncertainties in data acquisition, reducing biases in the astrophysical models that underly the probes, or taking into account observer-dependent variances in the parameters of the cosmological background model. 
Yet, early and late universe probes are often treated as independent, they live on different length scales, and require different perturbations to be subtracted.
Hence, fitting a flat Friedmann-Lemaître-Robertson-Walker cosmology to different probes at different cosmic epochs can yield different sets of cosmological parameter values. 
Tensions arise if these background fits and perturbing biases are not \emph{consistently} calibrated or synchronised with respect to each other. 
This consistent model-fitting calibration is lacking between the two $H_0$ values mentioned above, thus causing a tension.
As shown here, this interpretation resolves the $H_0$ tension, if 15\% of the matter-density parameter obtained from the fit to the cosmic microwave background, $\Omega_\mathrm{m}=0.315$, are assigned to decoupled perturbations yielding $\Omega_\mathrm{m}=0.267$ for the fit at redshifts of the supernova observations.
Existing theoretical analyses and data evaluations which support this solution are given.
}

\FullConference{%
  Corfu Summer Institute 2022 "School and Workshops on Elementary Particle Physics and Gravity",\\
  28 August - 1 October, 2022\\
  Corfu, Greece}

\begin{document}
\maketitle

\section{Introduction}
\label{sec:introduction}

Debates about the value of the Hubble constant $H_0$ have been ongoing almost since the introduction of Friedmann-Lemaître-Robertson-Walker (FLRW) cosmologies as viable models to interpret our observations. 
Within this framework of homogeneous, isotropic models, it is possible to introduce a global cosmic time and a corresponding, observable redshift $z$. 
The evolution of the universe is then described as the evolution of the cosmic expansion function over redshift $z$, with $H_0$ as the normalising scaling factor
\begin{equation}
H(z) = H_0 \; E(z)
\label{eq:H}
\end{equation}
with 
\begin{equation}
E(z) = \sqrt{\Omega_\mathrm{r} \left( 1+z\right)^4 + \Omega_\mathrm{m} \left(1+z \right)^3 + \Omega_\mathrm{k}  \left(1+z \right)^2 + \Omega_\Lambda} \;.
\end{equation}
The $\Omega_i$ denote the dimensionless density parameters for radiation, matter, curvature, and a cosmological constant/dark energy contribution to the total energy-momentum content from left to right.
They are the density parameters today at $z=0$ and $E(z)$ is normalised at $z=0$ such that 
\begin{equation}
\Omega_\mathrm{r} + \Omega_\mathrm{m}  + \Omega_\mathrm{k} + \Omega_\Lambda = 1
\label{eq:sum}
\end{equation}
and therefore $H(z=0) = H_0$.
All quantities involved are smooth fields evolving over cosmic time with $H_0$ setting the scale, see e.g.~\cite{bib:Ellis_book} for a general review on cosmological model building.

Contrary to expectations that an increasing amount and precision of observations and many independent methods to determine the parameters in \ref{eq:H} would converge to one precise set for probes at all $z$, tensions have aggravated. 
Focussing on the tension in $H_0$, currently the most recent value as obtained from 42 supernovae Ia at redshifts $z < 0.01$ by \cite{bib:Riess} lies around $H_0 = 73.04~\mbox{km/s/Mpc}$ with uncertainties reduced to 1.04~km/s/Mpc, while the value obtained by \cite{bib:Planck} fitting a $\Lambda$-Cold-Dark-Matter ($\Lambda$CDM) cosmology to the data of the cosmic microwave background (CMB) at $z \approx 1100$ yields $H_0=(67.27\pm0.60)~\mbox{km/s/Mpc}$. 
Numerous ideas have been put forward to resolve this $5$-$\sigma$ discrepancy, see the recent review paper by \cite{bib:DiValentino} and references therein for an encompassing overview. 
Among the suggestions were highly unconventional ideas like \cite{bib:Odintsov} proposing a model that the universe underwent a pressure singularity leading to an abrupt change in the cosmic evolution, which could even have had an observable impact on earth.  

\begin{figure}
\centering
\includegraphics[width=0.85\textwidth]{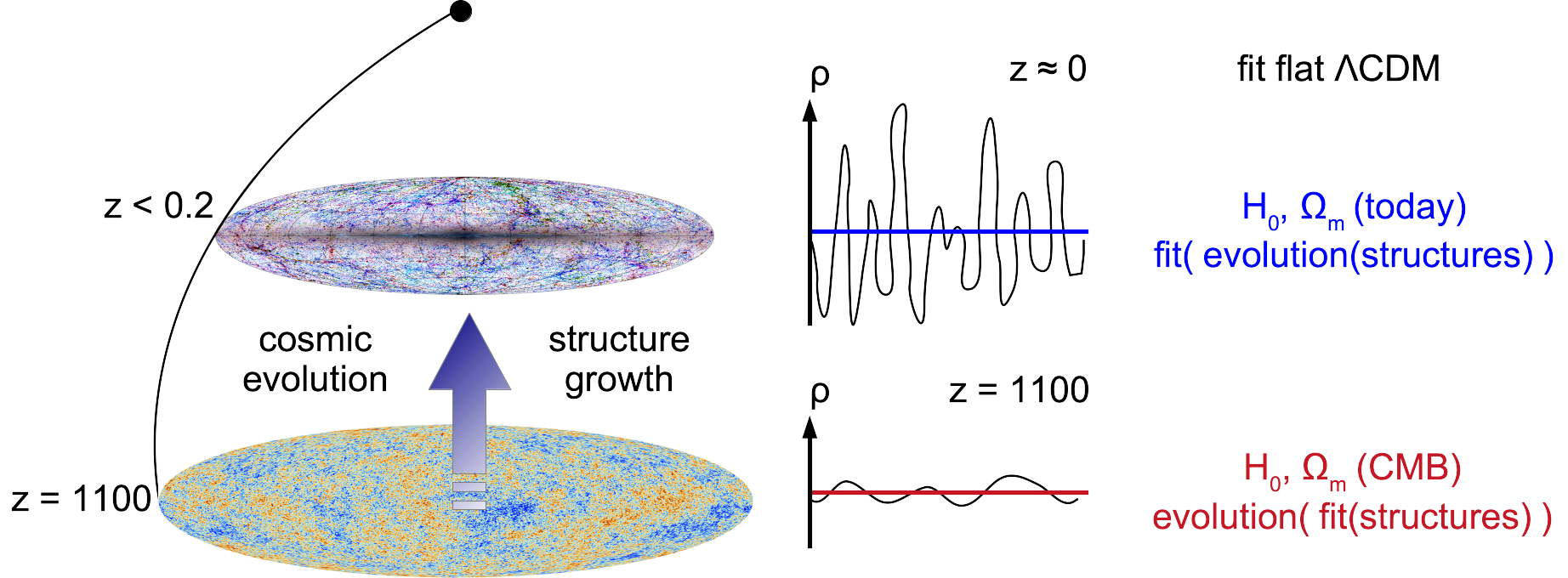}
\caption{Schematic of fitting an FLRW model to the CMB observations and data in the late universe. The challenge in the former is the foreground subtraction, while the latter suffers from finding the background among the strong non-linear perturbation level. For both fits to have the same cosmological parameters, one must check that the evolution of the cosmological model fit to the CMB actually describes the same cosmological background under the evolved non-linear structures in the late universe.}
\label{fig:fit}
\end{figure}

This work introduces yet another idea which does not seem to have been pursued yet. 
\cite{bib:Ellis} interpret the cosmological models with parameters introduced in \eqref{eq:H} as an idealised fitting function to observations in a universe with a higher degree of complexity.
Consequently, the imperfect fit yields a best-fitting background cosmology with a certain parameter set and perturbing residuals. 
Seen from the observational side, any observed signal consists of the emitted signal, a contribution due to the propagation in the symmetric background cosmology and a part caused by biasing propagation effects of perturbations\footnote{Biases caused by the data acquisition or insufficient knowledge of the emitted signal occur as well and have to be accounted for in addition to the argument made here.}. 
For example, consider the flux observed from a supernova explosion at high redshift which is redshifted while propagating through the background cosmology and (de-)magnified and deflected by biasing gravitational lensing effects. 
The difficulty to keep track of all these effects is detailed in \cite{bib:Clarkson}. 
On top of that, there can also be stray light from neighbouring light sources or dust extinction.
Data is often sparse and underdetermined, such that identifying the background cosmology from these signals requires a lot of additional model assumptions. 

Assuming a fully inhomogeneous, anisotropic real universe, fitting FLRW models to its different evolution stages can yield different parameter sets of $H_0$ and $\Omega_i$ (see e.~g.~\cite{bib:Clarkson2011} and \cite{bib:Schander} for overviews on back-reaction effects and \cite{bib:Heinesen} for more general fitting functions in geometries beyond FLRW), which is to be expected. 
But the same is also true for a \emph{statistically} homogeneous and isotropic, structured FLRW universe in which the split into smooth background fields and perturbing inhomogeneities is not performed consistently across all cosmic times for all observed signals. 
Care needs to be applied to consistently match the rather uniform, continuous matter density with small perturbations at the CMB onto a highly non-linear one at low redshifts.  
Separating a smooth background mass density from perturbations becomes increasingly intricate for growing non-linearities in decreasing space-time volumes when decreasing the redshift from $z\approx 1100$ to $z=0$ (see Fig.~\ref{fig:fit} for a visualisation).
The background $\Omega_\mathrm{m}$ from the fit at $z=1100$ thus may contain contributions that are considered decoupled perturbations in the FLRW fit at smaller redshifts. 
Such inconsistencies in calibrating the two FLRW fits at different redshifts with respect to each other can lead to deviations in $\Omega_\mathrm{m}$ and, as a consequence, $H_0$ compensating for that change because it is the normalisation in \eqref{eq:H} turning the matter density into a dimensionless $\Omega_\mathrm{m}$. 
A similar degeneracy between $\Omega_\Lambda$ and $H_0$ was noted in \cite{bib:Efstathiou} when analysing the degeneracies between cosmological parameters fitting observations of the CMB in the pre-Planck era to an FLRW model. 
Their treatment of $H_0$ as a ``secondary parameter'' is equal to its usage as a normalisation constant here setting \eqref{eq:sum}. 
As detailed in \cite{bib:Planck}, at $z \approx 1100$, the $\Omega_\mathrm{m}$-term dominates in \eqref{eq:H} and therefore, $H_0$ is not directly inferred from the data. 

Going to redshifts $z \approx 0$, \eqref{eq:H} can be expanded into
\begin{equation}
H(z) = H_0 \left( 1 + \frac32 \Omega_\mathrm{m} z + \mathcal{O}(z^2) \right)  \;.
\label{eq:H_taylor}
\end{equation}
For $z = 0.01$ and $\Omega_\mathrm{m}=0.3$, the second term in brackets is 0.0045. 
Hence, low redshifts observations like Cepheids and supernovae relevant for the local determination of $H_0$, as in \cite{bib:Riess}, are direct probes of $H_0$, but are not very sensitive to $\Omega_\mathrm{m}$.
As will be detailed below, choosing an embedding of our local neighbourhood into a global cosmology causes the $H_0$ tension because the embedding still lacks a synchronisation with the probes at the CMB epoch to make sure both fits identify the same background cosmology. 


\subsection{Outline}
\label{sec:outline}
Section~\ref{sec:linking} first introduces all definitions and requirements of the $H_0$-$\Omega_\mathrm{m}$ degeneracy responsible for the $H_0$ tension.
Embedding $H_0$ from \cite{bib:Riess} into a sample set of supernova, a full cosmological parameter set for a flat universe at late cosmic times is obtained. 
At the CMB, the parameters obtained by \cite{bib:Planck} are employed. 
Then, the portion of $\Omega_\mathrm{m}$ from the fit at the CMB is determined which has to deviate from the value obtained by the fit to the supernovae, such that the $H_0$ tension between \cite{bib:Planck} and \cite{bib:Riess} is resolved. 
Searching for supporting evidence that such a conversion from background matter density into perturbing matter density is a realistic solution, Section~\ref{sec:supporting_arguments} introduces relevant works and details possible reasons for the two misaligned FLRW fits leading to the $H_0$ tension.
Section~\ref{sec:conclusion} summarises all findings and highlights the importance of a consistently calibrated cosmological model for our understanding of cosmic structure evolution.
In all calculations, we only assume that the non-linearities in the late universe impede us from identifying the \emph{corresponding} background level as can be fitted at CMB time. 
Whether the universe deviates from a true FLRW geometry or back-reaction effects play a role is a different question because, as we will show, a mere misfit, also interpretable as a de-synchronisation between the early and late universe probes, already suffices to explain the tension.

\section{Linking the early- and late-universe fits}
\label{sec:linking}

\subsection{Prerequisites}
\label{sec:prerequisites}

Most of the observational probes rely on the measurement of electro-magnetic signals on the backward light-cone. 
These observed fluxes, surface brightnesses, or whole spectra are often measured in terms of absolute and apparent magnitudes, calibrated by reference objects. 
Applying these observational standards, cosmic distance measurements based on brightness are thus consistently set up across the entire cosmic evolution history. 
This brightness-based distance measure can be linked to a distance measure based on a cosmological model: 
Within a homogeneous and isotropic space-time geometry, a redshift is measured via spectroscopy and inserted into a distance measure containing a parametrised evolution function $H(z)$, as for instance given by \eqref{eq:H}. 
Yet, this model-based distance to the emitting source of the signal only accounts for the propagation of the signal through the background cosmology.
To match the measured brightness-based distance, changes to the model-based distance need to be implemented to account for interactions with perturbing structures on top of this background. 
Most common examples of such perturbing biases are gravitational lensing being independent of the observation wavelength or dust extinction which is wavelength-dependent. 

Restricting the class of fitting cosmologies to FLRW, it is possible to set up a single redshift variable $z$ across cosmic evolution history which represents the observable redshift and is inserted into the model-based distance measure. 
(Even if the real universe is not homogeneous and isotropic, it is still possible to set up a single $z$ and adjust the model-based distance measure or perturbing biases to account for the residuals, such that this prerequisite is only a consistent choice to fit an FLRW cosmology.
An analogous approach is made when incorporating the uncertainties of measured redshifts in the uncertainties of resulting distance estimates when determining supernova distances, see \cite{bib:Wagner5} for details on the principle of simplifying a fitting problem by keeping the independent variable as simple as possible.)
Then, to investigate the $H_0$ tension, two redshifts of relevance can be identified: the redshift of the CMB, $z_\mathrm{C}\approx1100$, and the redshift around which the supernovae in \cite{bib:Riess} are spread, $z_\mathrm{S} \ll z_\mathrm{C}$, which lies below $z=0.15$ with a mean redshift of $z=0.055$. 
As the volume spanned by the SNe in Cepheid hosts (roughly out to $z=0.011$) is too small to constrain an FLRW model (see Section~\ref{sec:introduction}), the sample is increased to include SNe in the Hubble flow ($0.023 < z < 0.15$). 
We will even extend the considerations such that the cosmic expansion function can be constrained by further supernovae like in \cite{bib:Wagner5}, out to $z_\mathrm{S} \approx 1$. 


On the modelling side, a flat universe is assumed, i.~e.~$\Omega_\mathrm{k}$ is set to 0. 
Since the entire range of considered redshifts lies in the matter-dominated part of the cosmic history, we fit FLRW models to both of them, containing $H_0$, $\Omega_\mathrm{m}$, and a cosmological constant term $\Omega_\Lambda = 1 - \Omega_\mathrm{m}$ (by means of \eqref{eq:sum}, also employed in \eqref{eq:H_taylor}).
To distinguish the resulting parameters of both fits, the parameters at $z_\mathrm{S}$ have a subscript~S, $H_\mathrm{0,S}$, $\Omega_\mathrm{m,S}$, $\Omega_\mathrm{\Lambda,S}$, while the ones at $z_\mathrm{C}$ obtain a~C, $H_\mathrm{0,C}$, $\Omega_\mathrm{m,C}$, $\Omega_\mathrm{\Lambda,C}$. 
To clarify the definitions, both parameter sets $\left( H_\mathrm{0,C}, \Omega_\mathrm{m,C} \right)$ and $\left( H_\mathrm{0,S}, \Omega_\mathrm{m,S} \right)$ are FLRW parameter sets at $z=0$ and their cosmic evolution follows the Friedmann equations. 
They are just determined from fits to observables in the real universe at the two different redshifts when the split into a smooth background cosmology and perturbations can be different for the reasons detailed in \cite{bib:Ellis} and the ones mentioned in Section~\ref{sec:supporting_arguments}. 


\subsection{Linking distance measures}
\label{sec:linking_distance}

The comoving distance based on the FLRW fit at CMB
\begin{equation}
D_\mathrm{C}(z) = \frac{c}{H_\mathrm{0,C}} \int \limits_{0}^z \frac{\mathrm{d}\tilde{z}}{\sqrt{\Omega_\mathrm{m,C} (1+\tilde{z})^{3} + \Omega_\mathrm{\Lambda,C}}}
\label{eq:dc}
\end{equation}
and the one based on the fit in the late universe 
\begin{equation}
D_\mathrm{S}(z) = \frac{c}{H_\mathrm{0,S}} \int \limits_{0}^z \frac{\mathrm{d}\tilde{z}}{\sqrt{\Omega_\mathrm{m,S} (1+\tilde{z})^{3} + \Omega_\mathrm{\Lambda,S}}}
\label{eq:ds}
 \end{equation}
should yield the same result for every $z$ in their common range of validity, i.e.
\begin{equation}
D_\mathrm{C}(z) \stackrel{!}{=} D_\mathrm{S}(z)
\label{eq:equation}
\end{equation}
if the model-based distances are both consistent with the brightness-based observational distances and perturbing biases have been taken into account\footnote{Observations employ angular diameter or luminosity distances, yet, their redshift-dependent prefactors cancel out in \eqref{eq:equation}.}. 
The range of validity can be restricted to the maximum redshift of the supernova sample, if one only assumes the cosmological parameter sets to be valid up to their outmost data point, which will be used below. 
Alternatively, assuming any probe can constrain the global cosmological parameters, \eqref{eq:equation} can be applied to any redshift up to the radiation dominated era, when $\Omega_\mathrm{r}$ needs to be included in $E(z)$. 

If all perturbing biases at $z_\mathrm{C}$ and $z_\mathrm{S}$ to the observed signals were the same, the two FLRW parameter sets would coincide. 
Yet, due to the dynamical evolution, the effective backgrounds and resulting perturbations of both fits are different a priori.
Hence, agreement in the cosmological parameters is only expected if the perturbation biases as a whole are matched consistently between the two fits, even if individual biasing effects at $z_\mathrm{S}$ and $z_\mathrm{C}$ are different.  

\begin{figure}
\centering
\includegraphics[width=0.48\textwidth]{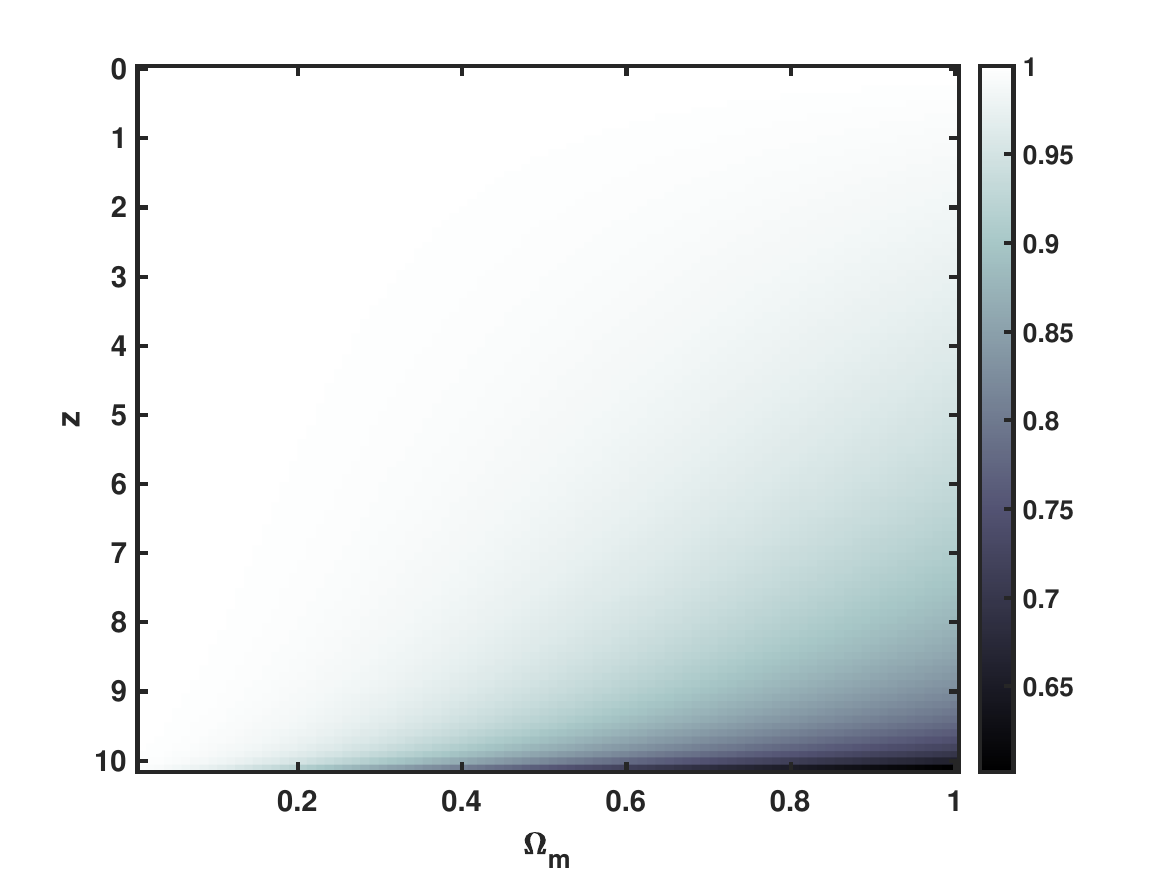}
\caption{Hypergeometric function value (indicated by colour) for $\Omega_\mathrm{m} \in \left[0.01, 1.00\right]$ ($\Omega_\Lambda = 1 - \Omega_\mathrm{m}$) and $z \in \left[0, 10\right]$, supporting the approximation that it is equal to 1 for cases to which it is applied in this work.}
\label{fig:hypergeom}
\end{figure}

Converting the redshift $z$ into the cosmic scale parameter $a$ to solve the integrals in \eqref{eq:dc} and \eqref{eq:ds} yields 
\begin{equation}
 \int \limits_{a}^1 \frac{\mathrm{d}\tilde{a}}{\tilde{a}^2 \sqrt{\Omega_\mathrm{m} \tilde{a}^{-3} + \Omega_\mathrm{\Lambda}}} = \Big[ 2 \sqrt{\frac{\tilde{a}}{\Omega_\mathrm{m}}} {_2}F_1 \left(\frac{1}{6},\frac{1}{2}; \frac{7}{6}; - \frac{\Omega_\mathrm{\Lambda} \tilde{a}^3}{\Omega_\mathrm{m}} \right) \Big]_{a}^1 \;.
 \label{eq:integral}
\end{equation}
Hence, the $\Lambda$-term only enters in the argument of the hypergeometric function ${_2}F_1$. 
As shown in Figure~\ref{fig:hypergeom}, the latter is slowly varying and always of the order of 1 for the physically reasonable range of $\Omega_i$ between 0 and 1 ($\Omega_\mathrm{m} \ne 0$), such that
\begin{equation}
{_2}F_1 \left(\frac{1}{6},\frac{1}{2}; \frac{7}{6}; - \frac{\Omega_\mathrm{\Lambda}\tilde{a}^3}{\Omega_\mathrm{m}} \right) \approx 1 \;.
\label{eq:approximation}
\end{equation}

For the ideal case that all perturbing biases have been correctly matched, inserting \eqref{eq:integral} with \eqref{eq:approximation} into \eqref{eq:equation} consequently yields
\begin{equation}
\frac{H_\mathrm{0,C}}{H_\mathrm{0,S}} \approx \sqrt{\frac{\Omega_\mathrm{m,S}}{\Omega_\mathrm{m,C}}} \;.
\label{eq:ratio}
\end{equation}
For $H_\mathrm{0,C}~=~67.27~\mbox{km/s/Mpc}$ from \cite{bib:Planck} and $H_\mathrm{0,S}~=~73.04~\mbox{km/s/Mpc}$ from \cite{bib:Riess} and using $\Omega_\mathrm{m,C}=0.315$ from \cite{bib:Planck}, \eqref{eq:ratio} can be solved for the matter-density parameter expected in the FLRW fit at $z_\mathrm{S}$ to fully resolve the $H_0$ tension, namely $\Omega_\mathrm{m,S}=0.267$.

Taking the FLRW fit at $z_\mathrm{C}$ as reference and the $H_0$ from \cite{bib:Riess} as fixed at $z_\mathrm{S}$, this result implies that the $H_0$ tension is resolved, if about 15\% of the matter assigned to the background density at $z_\mathrm{C}$ are decoupled from the background at $z_\mathrm{S}$ and are considered as perturbing biases. 
Hence, this can be interpreted that the local cosmic neighbourhood is an under-dense environment compared to the CMB background and can be considered a local void. 
Alternatively, this can be interpreted that we bias the cosmological matter density background in our local neighbourhood low and tend to assign more matter density to the perturbation level, meaning to the non-linear mass agglomerations.

The flat $\Lambda$CMD model could be relaxed to open up for further degrees of freedom or the contribution of neutrinos to all parameters could also be considered.
This only allows to redistribute the overall energy-density in more ways without changing the principle of the argument, as all $\Omega_i$ include a scaling by $H_0$ to become dimensionless parameters (see \eqref{eq:H}, \eqref{eq:sum}).

Since the observations to fit the FLRW model at $z_\mathrm{C}$ are all-sky, foreground biases are mostly at much lower redshifts and the observations probe an almost smooth matter density distribution with small perturbations, the fitting appears more well-defined at $z_\mathrm{C}$ compared to the one at $z_\mathrm{S}$, where a smooth background component of the matter is much more difficult to identify among highly non-linear, moving foreground perturbations. 
As Section~\ref{sec:supporting_arguments} shows, there are hints that perturbing biases are still not perfectly matched between the FLRW fits at $z_\mathrm{C}$ and $z_\mathrm{S}$. 
As a consequence, \eqref{eq:ratio} is most likely a maximum discrepancy between the two FLRW fits because inconsistently matched perturbing biases (which can be incorporated as a multiplicative factor $b$ on the right-hand side of the equation) alleviate the deviation between the $\Omega_\mathrm{m}$. 
Overall, given the estimated amount of matter that needs to be redistributed even for an ideal matching of perturbation effects, it is possible to resolve the $H_0$ tension by this approach, most likely without the need to involve further cosmological parameters, as detailed in the following Section~\ref{sec:supporting_arguments}.

\section{Supporting arguments in observations}
\label{sec:supporting_arguments}

From the theory side, the impact of fitting a smooth function to a clumpy (Swiss-cheese-model) universe based on supernova observations was investigated in \cite{bib:Fleury}. 
It was found that $\Omega_\mathrm{m}$ can indeed be underestimated in the presence of non-uniformities, such that the estimated deviations between $\Omega_\mathrm{m,C}$ and $\Omega_\mathrm{m,S}$ in Section~\ref{sec:linking_distance} can be explained.
Indeed, several FLRW fits to supernova data, e.~g.~\cite{bib:Astier, bib:Betoule, bib:Scolnic, bib:Wagner5, bib:Colgain} yielded  $\Omega_\mathrm{m,S} < \Omega_\mathrm{m,C}$, supporting the theoretical hypothesis. 
\cite{bib:Umeh} also supports the hypothesis based on a more general calculation showing that distance measures at $z < 0.001$ are underestimated by 4-12\% due to inhomogeneous structures on top of the cosmological background which was fixed by the values of \cite{bib:Planck}. 

In \cite{bib:Camarena}, a Lemaître-Tolman-Bondi model is fitted to the Pantheon sample of supernovae, once restricting the data to $z_\mathrm{S} < 0.15$ and once using the full sample $z_\mathrm{S} < 2.3$. 
For the low-$z$ sample, evidence for an underdensity is found also for this cosmological model taking into account a local inhomogeneity beyond FLRW. 
Even though the full supernova sample does not continue to support the existence of a local void beyond approximately 100~Mpc in this analysis, the authors cannot exclude its existence because their approach does not account for anisotropies, which were taken into account in complementary evaluations of local observations supporting such an underdensity, as e.g.~summarised in \cite{bib:Haslbauer}. 

Independently, evidence for a local underdensity is reproduced by galaxy-cluster counts at low redshift, see \cite{bib:Boehringer} and references therein.
\cite{bib:Boehringer} also discuss the $H_0$ tension based on the correlation between $\Omega_\mathrm{m}$ and $H_0$, concluding that an under-dense matter-density at $z_\mathrm{S}$ compared to an $\Omega_\mathrm{m}$ obtained by averaging over a larger space-time volume implies an increase in $H_0$ compared to the $H_0$ parameter of the larger space-time volume.
Corresponding theoretical investigations of cluster-count statistics in \cite{bib:Heinesen2} analyse the impact of the survey selection function on identifying inhomogeneities in the matter-density distribution and corroborate that a bias towards homogeneity can occur as a result of a selection function overcorrecting for inhomogeneous sampling. 

 \cite{bib:Rameez} discovered inconsistencies between the JLA and the Pantheon supernova data set at low redshifts.
 The changes in the preprocessing of the two data sets are supposed to be bias corrections due to the motion of the heliocentric frame with respect to the CMB. 
 The fact that these bias corrections can shift $H_0$ from $68~\mbox{km/s/Mpc}$ (JLA) to $72~\mbox{km/s/Mpc}$ (Pantheon) and that it is debated whether or not they should be applied clearly shows the difficulty to separate the smooth background from perturbations in the highly non-linear regime.  
 
Contrary to supernovae, baryon acoustic oscillations (BAOs) at low redshifts are not difficult to be consistently calibrated with respect to the CMB because the sound horizon is a feature which is easy to track in correlation functions or power spectra, particularly because its scale lies beyond the range in which non-linear structures dominate the matter density. 
Retrieving $H_0$ by calibrating a supernova set using BAOs instead of local Cepheids, \cite{bib:Macaulay} find $H_0~=~(65.7\pm2.4)~\mbox{km/s/Mpc}$, well in agreement and even lower than the one from \cite{bib:Planck}. 
They conclude that the consistency of their value to $H_0$ by \cite{bib:Planck} is a mere consistency of all probes involved tracing the same cosmological model, corroborating the statements in Section~\ref{sec:linking} because their reference is the fit to the rather homogeneous and isotropic energy-momentum content at $z_\mathrm{C}$. 
As shown in \cite{bib:Planck}, Section~5.4.~and references therein, this consistency is maintained for other datasets with BAOs.
 
As a notable further finding, \cite{bib:Betoule} investigate the impact of a recalibration, including a readjustment of zero-points for two filters.
The recalibration led to a shift of the matter density parameter from $\Omega_\mathrm{m,S}=0.230$ to $\Omega_\mathrm{m,S}=0.291$.
Thus, a recalibration of zero-points of the observed brightness can also lead to an increased consistency between the FLRW fits at $z_\mathrm{C}$ and $z_\mathrm{S}$. 
This is an example of an inconsistent matching between perturbing biases before equating the FLRW-model-based distances in \eqref{eq:equation}. 


The clearest hint to support the hypothesis that the tension is a mere inconsistent comparison of two FLRW fits, comes from the strong-lensing-inferred $H_0$ using elliptical galaxies as gravitational lenses.
As detailed in \cite{bib:Wagner4, bib:Wagner6}, the strong gravitational lensing formalism bears mathematical degeneracies that physically amount to the freedom to partition the total line-of-sight mass density into a lens, potentially perturbing masses, and a smooth background cosmological embedding.  
Fixing the partitioning usually requires complementary data like velocity dispersions along the line of sight and spectroscopic analyses of the immediate surroundings to constrain the invisible distribution of dark matter well enough to fix the lensing mass distribution and line-of-sight perturbing masses on top of an embedding smooth background.
\cite{bib:Birrer} show that varying constraints on the mass density profile of the gravitational lens can yield values of $H_0$ between $67.4~\mbox{km/s/Mpc}$ and $74.5~\mbox{km/s/Mpc}$.
The resulting $H_0$ value is anchored in observables retrieved in a highly non-uniform mass distribution (see \cite{bib:Ding} for a recent benchmark test on various approaches), which gives an explanation for this variance. 
Since this way of fixing the partitioning is completely unrelated to the one at $z_\mathrm{C}$ and happening in a highly non-linear regime, a coincidence of $H_0$ values implies that the overall perturbing biases (including selection effects) are finally brought to agreement to those at $z_\mathrm{C}$.

Yet, the strong-lensing $H_0$ values are mostly higher than $H_0$ at $z_\mathrm{C}$ and consistently follow the trend set by other probes in very non-uniform environments, as e.~g.~detailed in \cite{bib:Wong}, supported by the findings of \cite{bib:Fleury}. 
The complementarity of this probe and its independence from the ones in the early universe, which was first considered an advantage, is thus exactly the issue causing the lack of synchronisation required to resolve the tension between $z_\mathrm{S}$ and $z_\mathrm{C}$. 



A related $H_0$-$\Omega_\mathrm{m}$-degeneracy was recently brought forward by \cite{bib:Colgain2022}.
The authors systematically investigated the changes in the cosmological parameters when fitting a $\Lambda$CDM model to quasar, supernova, and BAO observations for samples of different effective redshift, i.e.~restricting data sets to different, limited ranges of redshift.
They find that the fits yield lower $H_0$ and higher $\Omega_\mathrm{m}$ when increasing the effective redshift. 
To explain this trend, the authors argue that the most relevant terms in the model fit change with increasing effective redshift, leading to a change in the best-fit cosmological parameter values.
For low redshifts around $z=0$, the leading degree of freedom in $\Lambda$CDM is $H_0$ and $\Omega_\mathrm{m}$ is hardly constrained in our local neighbourhood. 
Increasing the effective redshift of the sample, the leading term becomes proportional to $H_0 \sqrt{\Omega_\mathrm{m}}$, thus anti-correlating $H_0$ and $\Omega_\mathrm{m}$ as in \eqref{eq:ratio}. 
If their result persists with increasing statistical significance, the suggestion to use BAOs to resolve the $H_0$ tension may not be easily feasible because the authors also find a trend of increasing $\Omega_\mathrm{m}$ for this cosmic probe at $z < 1$, which requires follow-up investigations to understand its cause.

On the whole, it is very likely that an FLRW fit to a clumpy universe at $z_\mathrm{S}$ complicates a consistent matching to an FLRW fit at $z_\mathrm{C}$ if no comparable reference between the observables at these redshifts is employed, like, for instance the sound horizon. 
Since the latter can be traced over a wide redshift range already, BAOs are ideal observables to establish consistencies between FLRW fits in this range.  
The above arguments indicate that the local matter density is most likely biased low, resulting in different parameter values for $H_0$ and $\Omega_\mathrm{m}$ at $z_\mathrm{C}$ and $z_\mathrm{S}$, causing the $H_0$ tension.
One would expect that consistency is also facilitated for probes beyond the scale of homogeneity, as the BAOs suggest.
Yet, whether such a scale actually exists remains to be demonstrated (see \cite{bib:Maartens} for an overview and e.~g.~\cite{bib:Heinesen2, bib:Goncalves} and references therein for recent advancements), which can be questioned by the detection of ever increasing inhomogeneous structures at high redshifts, see e.~g.~\cite{bib:Lopez} for a recent example.


\section{Conclusion}
\label{sec:conclusion}

Assuming the 5-$\sigma$ tension in the values of the Hubble constant $H_0$ determined by \cite{bib:Planck} and \cite{bib:Riess} cannot be resolved by the numerous ideas summarised e.~g.~in \cite{bib:DiValentino}, this work puts forward another way of resolving this $H_0$ tension. 
Based on the viewpoint detailed in \cite{bib:Ellis}, the observations at the CMB are assumed to determine one set of cosmological parameters of a flat FLRW geometry at redshift $z=1100$. 
In the same manner, the observations of supernovae constrain a second set of FLRW cosmological parameters at low redshift. 
Only if both fits consistently define the split of the overall energy density of the cosmos into background homogeneous fields and perturbing inhomogeneities, the two parameter sets will coincide.
Most notably, the local $H_0$ value as determined in \cite{bib:Riess} requires an embedding in an overall cosmology in the first place as the observational probes do not reach out far enough for the cosmological parameter fit to be sensitive to $\Omega_\mathrm{m}$ (estimated in Section~\ref{sec:introduction}). 

After embedding the local $H_0$ value into a cosmology constrained by supernova data up to redshifts of order 1, Section~\ref{sec:linking} detailed a way to perform such a matching of FLRW fits for observational data that are consistently calibrated by observational reference standards.
Thus, the $H_0$ tension is no tension in a single cosmological parameter but rather a consequence of an inconsistency between two entire sets of cosmological parameter values in which $H_0$ as a normalisation constant sets the scale for all parameters, see \eqref{eq:H}.

As supported by the arguments put forward in Section~\ref{sec:supporting_arguments}, the $H_0$ tension can be interpreted as a mere offset in the reference amplitude for the smooth cosmic matter density parameter between the CMB and in the late universe.
This offset seems most likely to be caused by the difficulty to consistently retrieve the high-redshift reference amplitude at low redshifts in a highly inhomogeneous local environment. 

This explanation also implies that the $H_0$ tension can be resolved as a mere consistent model fitting of observables at a given resolution and precision to a statistically homogeneous and isotropic universe (or even to an a priori arbitrarily complex universe). 
As such, it does not necessarily yield any information about the real geometry of the universe.
Yet, resolving the $H_0$ tension by setting up a consistent split into background and foreground for all astrophysical phenomena allows for a consistent reconstruction of cosmic structure evolution across its entire observable stages.
It bridges the gap between the large-scale, all-sky and small-scale, local observations and models. 
The resulting calibration of all cosmic probes to one common, synchronised framework will thus provide the grounds to gain a deeper understanding of cosmic structure evolution than we have now.
A mere convergence of independent probes to the same cosmological parameter values may not accomplish this task. 
Whether or not FLRW models provide the appropriate framework to achieve consistency in the most convenient way for the increasingly detailed data in upcoming surveys remains an open question to be further investigated alongside.

\section{Acknowledgements}

I cordially thank George Ellis for his timelessly topical papers, the workshop \href{https://sites.google.com/apctp.org/cosmoprinciple}{``A Discussion on the Cosmological Principle"} and the \href{https://www.lhc-epistemologie.uni-wuppertal.de/news/news/discussion-group.html}{``Peebles Fan Club"} philosophy-of-science seminar from the Lichtenberg Group of History and Philosophy of Physics at the University of Bonn for all the inspiring insights that contributed to write this work. 
In addition I am grateful for fruitful discussions on the draft with Chris Clarkson, Eoin Ó Colgáin, Richard Griffiths, Jori Liesenborgs, and David Wiltshire.


%

\bibliographystyle{agsm}
\bibliography{H0} 

\end{document}